\documentclass{article}
\usepackage{etex}

\usepackage[margin=1in]{geometry}
\usepackage{amsmath,amssymb,amsthm}
\usepackage{gensymb}

\usepackage{graphicx}

\usepackage{placeins}
\usepackage{setspace}

\usepackage{algorithm}
\usepackage[noend]{algpseudocode}
\algrenewcommand{\algorithmiccomment}[1]{\hskip\algorithmicindent \# #1}

\usepackage{caption}
\usepackage{subcaption}

\usepackage{multirow}

\theoremstyle{plain}
\newtheorem{theorem}{Theorem}[section]
\newtheorem{definition}[theorem]{Definition}

\usepackage{hyperref}	
\hypersetup{colorlinks,breaklinks,
            linkcolor=[rgb]{0,0.5,0.5},
						citecolor=[rgb]{0,0.5,0.5},
						pdfprintscaling=None}
\usepackage[all]{hypcap}
\usepackage[nameinlink,noabbrev]{cleveref}

\Crefname{ALC@unique}{Line}{Lines}
\Crefname{thm}{Theorem}{Theorems}
\crefname{thm}{theorem}{theorems}

\DeclareMathOperator{\Span}{span}

\title{Reorganizing topologies of Steiner trees to accelerate their elimination}

\author{
  Aymeric Grodet\thanks{Corresponding author} \thanks{E-mail addresses: aymeric.grodet@gmail.com,
    tsuchiya@math.sci.ehime-u.ac.jp.}
  \and
  Takuya Tsuchiya\footnotemark[2]
}
\date{Graduate School of Science and Engineering, \\Ehime University, 2-5, Bunkyo-cho, Matsuyama, Japan}

\makeatletter
\providecommand\theHALG@line{\thealgorithm.\arabic{ALG@line}}
\makeatother

\begin{document}
\maketitle

\begin{abstract}
We describe a technique to reorganize topologies of Steiner trees by exchanging neighbors of adjacent Steiner points. 
We explain how to use the systematic way of building trees, and therefore topologies, to find the correct topology after nodes have been exchanged. 
Topology reorganizations can be inserted into the enumeration scheme commonly used by exact algorithms for the Euclidean Steiner tree problem in $d$-space, providing a method of improvement different than the usual approaches. 
As an example, we show how topology reorganizations can be used to dynamically change the exploration of the usual branch-and-bound tree when two Steiner points collide during the optimization process. 
We also turn our attention to the erroneous use of a pre-optimization lower bound in the original algorithm and give an example to confirm its usage is incorrect.
In order to provide numerical results on correct solutions, we use planar equilateral points to quickly compute this lower bound, even in dimensions higher than two. 
Finally, we describe planar twin trees, identical trees yielded by different topologies, whose generalization to higher dimensions could open a new way of building Steiner trees.
\end{abstract}



\section{Introduction}

The Euclidean Steiner tree problem in $\mathbb{R}^{d}$ is an NP-hard optimization problem \cite{Garey} that can be defined as follows: given a set of points in $\mathbb{R}^{d}$, find the shortest tree (in terms of Euclidean length) that connects this set of points, using some additional points to construct the tree if necessary. 
For consistency, we refer to the original points as \emph{regular points}, even though they are often referred to as \emph{terminal nodes} or \emph{terminal points}. 
The additional nodes of the tree are called \emph{Steiner points}. 
The solution to this problem is called a Steiner minimal tree. 

\begin{figure}
\centering
\includegraphics[scale=0.7]{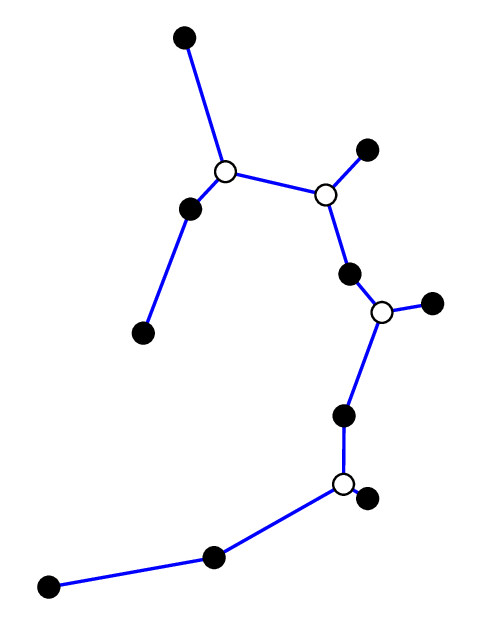}
\caption{Steiner minimal tree for a set of 10 regular points (black). The white dots are the Steiner points\label{Fig1}.}
\end{figure}

If a Steiner tree is defined as a tree with $N$ regular points to which we can add $K$ Steiner points, some of its well-known and useful properties \cite{Gilbert} are:
\begin{itemize}
\item{Every angle between incident edges must be greater than or equal to 120\degree.}
\item{Every vertex has a degree between 1 and 3, with the special case of the Steiner points which have three neighbors. Therefore, the edges incident to a Steiner point form angles of 120\degree. }
\item{The number $K$ of Steiner points is at most $N - 2$.}
\end{itemize} 
Readers may refer to \cite{ref4} to obtain more information about this problem, its properties and generalizations. Figure~\ref{Fig1} shows the Steiner minimal tree for the set of black regular points. The white points are the Steiner points. Here, $N=10$ and $K=4$. 

Smith \cite{Smith} described a \mbox{branch-and-bound} algorithm for the Euclidean Steiner tree problem in $d$-space. His algorithm relies mainly on two points: a specific enumeration scheme for all full Steiner topologies, and an effective optimization to find the minimum length of a Steiner tree. We refer to the combinatorial structure of a Steiner tree as its \emph{topology}, that is, how the points are connected to each other. 
A \emph{full topology} is a topology with the maximum of $N - 2$ Steiner points. If certain edges have shrunk so as to have zero length (for example, if a Steiner point is in the same position as a regular point), we say that the topology is \emph{degenerate}.
Because the positions of the regular points are fixed, the whole problem can be thought of as finding the topology that generates the Steiner minimal tree, and the position of the Steiner points that minimize the length of that tree.

We recall that \mbox{branch-and-bound} algorithms are a useful paradigm for combinatorial optimization problems (and others) that consists of a systematic enumeration of candidate solutions, those candidates forming a rooted tree. 
The algorithm explores branches of the tree, first checking the candidate branch solutions against a bound of the optimal solution. 
If the candidate solution is worse than the bound, this branch is cut from the tree. 

\begin{figure}
\centering
\includegraphics[scale=0.8]{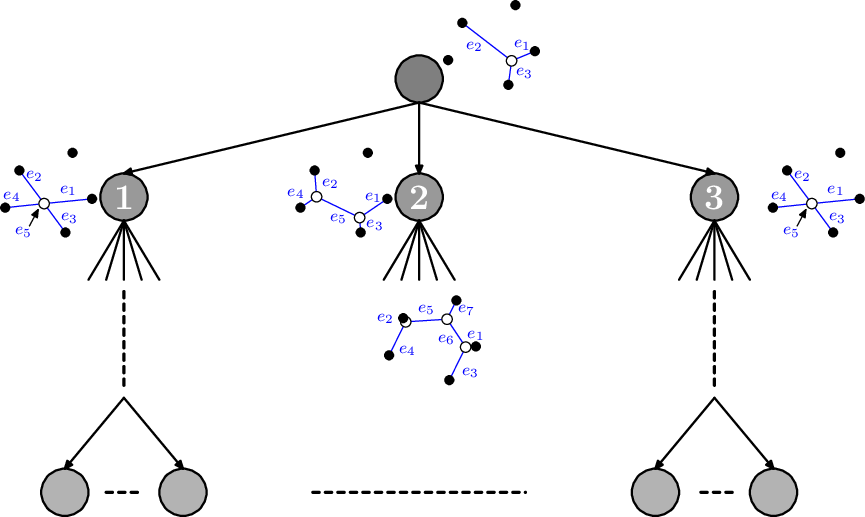} 
\caption{\mbox{Branch-and-bound} tree. The root is the unique tree with the first three regular points. The number of children of any node is equal to the number of edges in the Steiner tree corresponding to that node. Here, we have three Steiner trees with four regular points, created by merging $e_1$, $e_2$ and $e_3$, from left to right, as well as one fathered by the second topology with four points, by merging with $e_5$. Note that two of the trees look alike. This is due to two Steiner points located in the same position, we cannot see $e_5$ because it is joining them.   \label{bbtree}}
\end{figure}

For our problem, nodes in the branch-and-bound tree correspond to full topologies, and each child is created by a \emph{merging process} of a new regular point with one edge of the current Steiner tree. 
This process consists of deleting the selected edge and joining its two extremities and the regular point we want to add to a new Steiner point. 
Each merging process therefore adds two edges and one Steiner point to the current Steiner tree. 
This operation will be explained in more details in \cref{sec:topreorga}.
The root of the \mbox{branch-and-bound} tree is the unique full topology for three regular points, meaning the three are connected to one Steiner point. 
An immediate result is that the number of full topologies for $N$ regular points, that is, the number of nodes on the level $(N-2)$ of the branch and bound tree, is $1\times3\times5\times...\times(2N - 5)$. 
Therefore, one of the main challenges of this algorithm is to efficiently prune the branch-and-bound tree to minimize the number of explored topologies. 
\Cref{bbtree} shows how the topologies are organized in the branch-and-bound tree. 
The Steiner minimal tree is one of the leaves, so the length of any leaf gives an upper bound on its size. 
Note that adding one point to a Steiner tree (going down one level in the \mbox{branch-and-bound} tree) cannot decrease its length. In the extreme case, if the new regular point is exactly on the edge we are merging, the length will stay the same. 
In any case, it cannot decrease. Therefore, any upper bound for the Steiner minimal tree is also a bound for the other Steiner trees. 
It then seems logical that, if we can find better upper bounds faster, we can prune the \mbox{branch-and-bound} tree more.

A few papers proposed improvements to Smith's algorithm. 
In \cite{Fampa}, the authors used a conic formulation for the problem of locating the Steiner points in order to compute rigorous lower bounds, allowing a strong branching strategy. 
The substantial gain in pruned nodes was however compensated by the time require to compute the heavy procedure. 
Laarhoven and Anstreicher\cite{Laarhoven} presentend geometric conditions to create a fathoming criterion not based on lower bounds. 
In \cite{Fonseca}, the authors proposed improvements to Smith's algorithm and a new branch enumeration algorithm in a sense close to the approach used by GeoSteiner, the powerful solver for planar problems. 
An overview of exact algorithms in $d$-space is given in \cite{review}. 
The three common approaches to improve Smith's algorithm are the choice of a better fathoming criterion, to decide the order in which regular nodes are inserted, and the computation of the optimization process.

In this paper, we present a method to reorganize topologies by exchanging neighbors of adjacent Steiner points. 
This idea may at first glance seem unrelated to the points discussed in the previous paragraph. 
However, we describe a way to incorporate it into any algorithm cited above since it can be used to improve the enumeration scheme \emph{during} the optimization process, a strategy that the authors have not yet seen in the literature. 
Considering that the current algorithms are based on Smith's, we will give some computational results that show that using topology reorganizations in the original algorithm does reduce the number of topologies explored as well as the execution time. 
However, Smith made an error in the program he published whereby the error figure for a tree is used as a pre-optimization pruning method. 
Although it is already known that erroneous fathoming can occur in the original algorithm, we will give in \cref{appendixA} an example to illustrate how the use of the error figure can lead to incorrect solutions. 
Since this paper mainly focuses on relation between topologies, we will simply correct Smith's use of lower bound before the optimization process by using the idea of \emph{toroidal image} \cite{ref2}. 
Toroidal images use the set of equilateral points of two regular points to create the lower bound. 
This set becomes more complicated as we extend to higher dimensions, but in \cref{sec:pruning} we will use planar equilateral points to compute a lower bound quickly, and we refer the reader to the papers cited above for more rigorous lower bounds.

In \cref{sec:results}, we will see in what measure topology reorganizations can improve the original enumeration scheme. 
The results indicate that it is always worth using topology reorganizations during the computation of the optimization process of the original algorithm.

If topologies $T_1,...,T_m$ are generated from a topology $T$, we call $T_1,...,T_m$ the \emph{child topologies} of the \emph{parent topology} $T$. 
Note that a node in the \mbox{branch-and-bound} tree refers to a unique topology. 
Hence, both \emph{node} and \emph{topology} describe the same Steiner tree, and we will therefore use the two terms interchangeably in this paper. 
However, a unique topology does not necessarily refer to a unique Steiner tree. 
In \cref{sec:twintrees}, we will discuss what we call \emph{twin trees}, when two or more topologies lead to the same Steiner tree.

\section{A dynamic enumeration scheme}
\label{sec:scheme}
\subsection{The basic idea}
As briefly described in the previous section, the creation of a Steiner tree is a step-by-step process starting from the unique tree with three regular points. 
At each step, one edge is selected to be deleted and its extremities are linked to a new Steiner point, which is itself linked to the next regular point. 
Following this method of building Steiner trees, we can express a topology as the list of edges that have been removed to introduce the Steiner points. 
\begin{definition}
A topology is defined as $T=(t_1,...,t_j), j\leq N-3$ with $t_i$ being the index of the edge that has been deleted to introduce the $i$-th Steiner point $S_i$. 
\end{definition}
Note that $S_0$, the Steiner point introduced for the unique root topology with three regular points, does not need to be indicated in a topology. 
In the case of \cref{bbtree}, the topologies represented by the trees with four regular points are, from left to right, $(1)$, $(2)$, $(3)$ and the one with five points is $(2,5)$. 

A useful observation, already made by Smith, is that adding a new point to a Steiner tree by this process cannot decrease the length of the tree. 
Therefore, a good approach is to explore shorter topologies first, in the initial stages of the search, because they have a better chance of leading to shorter children, eventually providing an upper bound of higher quality (leaf of the branch-and-bound tree). 
Such a descent to the first upper bound is however a simplistic one, as one can use an initial bound (other than $\infty$) to increase the number of nodes fathomed. 
Once the upper bound is found, we want to grow partial solutions in such a way as they are likely to be pruned as quickly as possible (see for example \cite{Fonseca} for more details on these two points).
The enumeration scheme proposed by Smith is described by \cref{SES}. 

\begin{algorithm}
\renewcommand{\algorithmicrequire}{\textbf{Input:}}
\renewcommand{\algorithmicensure}{\textbf{Output:}}

\caption{Smith's enumeration scheme \protect\cite{Smith}} 
\label{SES}

\begin{algorithmic}[1]
\Require Set of regular points.
\Statex\hrulefill
\begin{spacing}{1.2}
\Statex \textbf{Step 0} $p \gets$ root node, $N = 3$
\State $l \gets \infty$

\Repeat
	\Statex \Comment \textbf{Step 1:}
	\State Compute the Steiner tree of all the child topologies of the current node $p$. 
	\State (if minimal length of a child tree $> l$, eliminate its subtree.)
	\Statex \Comment \textbf{Step 2} Go one level down:

	\If {all children are eliminated}
		\State Goto Step 3
	\Else
		\If {the child nodes are not leaves}
			\State Order the child topologies.
			\State $p \gets$ smallest child
			
		\Else 
			\ForAll{child topologies $p$}
				\If{minimal length of $p$ $< l$}
				\State $l \gets$ minimal length of $p$ 
				\State $S \gets p$
				\EndIf
			\EndFor
			\State Goto Step 3
		\EndIf
	\EndIf
\Until forever

\Statex \Comment \textbf{Step 3} Backtracking:

\If{all parent topologies have been explored} 
	\If{parent topology = root node}
		\State \textbf{exit}
	\Else
		\State Backtrack to one level higher (Goto Step 3).
	\EndIf
\Else
\State $p \gets$ smallest parent topology not yet explored
\State Goto Step 1
\EndIf

\end{spacing}
\hrulefill
\Ensure Steiner minimal tree $S$ for the input set.

\end{algorithmic}
\end{algorithm}

He proved the following theorem for his optimization process for locating the Steiner points for a given topology: 
\begin{theorem}[Smith, \cite{Smith}]
\label{thmSmith}
From all initial choices of Steiner point coordinates, except for a set of measure zero, the iteration converges to the unique optimum Steiner point coordinates that minimize the total tree length. This convergence happens in such a way that the sequence of tree lengths is monotonically decreasing.
\end{theorem}

In other words, the optimization process is a succession of generally convergent iterations to find the optimum coordinates of the Steiner points. One iteration consists of simultaneous updates of the current positions of all the Steiner points. Each of these points takes the position of the optimum Steiner point of its three neighbors. During this process, it often happens that two Steiner points reach the same position, or become close enough to effectively have the same position. 
We say that these points \emph{collide}. 
An example of this situation can be seen in topologies $(1)$ and $(3)$ in \cref{bbtree}. 
This is where \mbox{\emph{topology reorganizations}}, as described in the following section, can become useful. 
This idea comes from Chapeau-Blondeau, Janez, and Ferrier \cite{ref3}, who presented a relaxation scheme to locally optimize the length of a given Steiner tree, starting from the minimal spanning tree. 
In that scheme, there is a possibility of reorganizing the connections between two neighboring Steiner points. 
We are looking after an exact algorithm so we apply their idea of an \mbox{\emph{interaction process}} in the enumeration scheme. 
When a collision happens, we potentially operate a topology reorganization to dynamically jump to another, hopefully better, branch of the branch-and-bound tree.

\begin{figure}
\begin{center}
\includegraphics[scale=0.7]{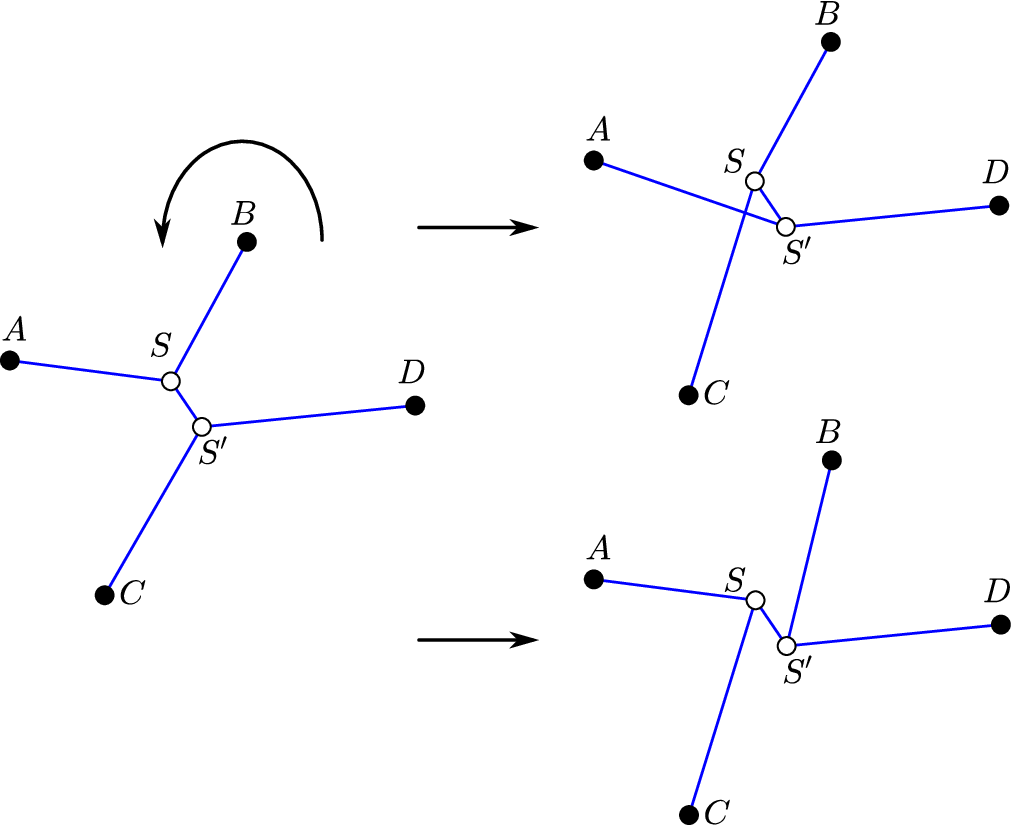} 
\end{center}
\caption{Possible reorganizations if $S$ and $S'$ collide: staying in the same situation, or $S'$ keeps $D$ and exchange $C$ with $A$ (top), or with $B$ (bottom).\label{Fig2}}
\end{figure}

Let us consider two Steiner points $S$ and $S'$ with respective neighbors $\{A,B,S'\}$ and $\{C,D,S\}$. 
If, at any time during the optimization process, $S'$ comes within a small enough collision distance of $S$, $S'$ considers the eventuality of exchanging one neighbor with $S$. 
There are only three possibilities to consider for the new neighborhood of $S'$: $\{A,D,S\}$, $\{B,D,S'\}$ and the current one $\{C,D,S\}$, the other ones being permutations of these. 
For computational reasons that will be explained in the following, the choice of the neighbors to exchange is important. 
\Cref{Fig2} shows an example of the possible reorganizations when $S$ and $S'$ collide. 

Note that it was proved in \cite[paragraph 8.4]{Gilbert} that the distance between two Steiner points must be at least $(\sqrt{3}-1)L_0$, where $L_0$ is the length of the shortest edge joining these two points to one of their other neighbor. However, we did not investigate if using this fact was computationally better than directly specifying a collision distance that we consider small enough to say that the two Steiner points collide.

\subsection{Topology reorganization}
\label{sec:topreorga}
Exchanging two neighbors of two adjacent Steiner points is simple, but it is not enough. 
If we want to use reorganizations in the exploration of the \mbox{branch-and-bound} tree, that is during the enumeration all the (full) topologies, we need to know which topology the reorganized structure corresponds to, or we may explore it and its descendants several times.
The original algorithm proposed by Smith used a stack to manage the child topologies and explore them in order. 
That way, there was no need to implement a proper tree structure to explore the \mbox{branch-and-bound} tree. 
However, we are now jumping from one node to others of same depth. 
Such a data structure would consequently ask a lot of resources to be managed, resources that would strongly impact the speed of the computation. 
We therefore made the choice to implement a tree structure to more easily manage which topologies we already optimized, explored, or pruned. 
The exploration of such a structure naturally calls for a recursive algorithm equipped with control instructions to avoid doing the same work several times.

\begin{figure}
	\centering
	\subcaptionbox{The root topology.\label{fig:topo0}}
		[0.3\textwidth]{\includegraphics[scale=0.8]{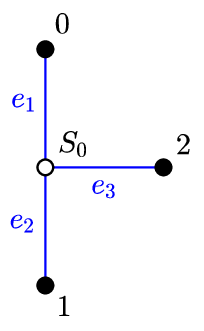}} \quad
	\subcaptionbox{Topology $(3)$. $g(2) < g(S_0)$ so $e_3 = e_{t_1}$ now links $2$ to $S_1$. \label{fig:topo3}}
		[0.3\textwidth]{\includegraphics[scale=0.8]{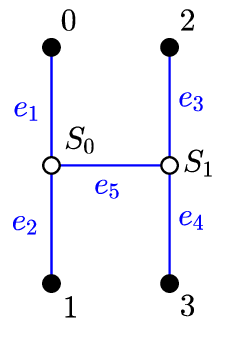}} \quad
	\subcaptionbox{Topology $(3,5)$.  $g(S_0) < g(S_1)$ so $e_5 = e_{t_2}$ now links $S_0$ to $S_2$.\label{fig:topo35}}
		[0.3\textwidth]{\includegraphics[scale=0.8]{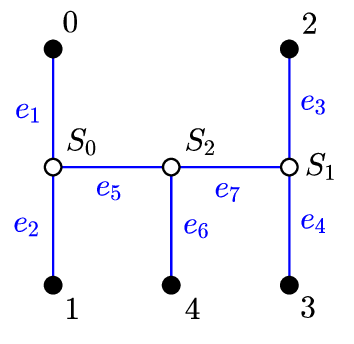}}
	\caption{Building a Steiner tree schematically.}\label{fig:topo}
\end{figure}

\begin{figure}
	\centering
	\subcaptionbox{The topology $(3,5,3)$. Inserting $S_4$ will create the topology $T_p=(3,5,3,9)$.\label{fig:reorga3}}
		[.4\textwidth]{\includegraphics[width=.4\textwidth]{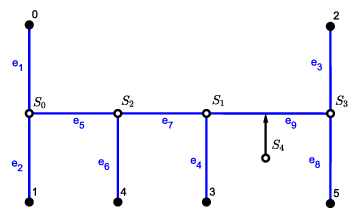}} \quad
	\subcaptionbox{When $S_4$ is inserted, it is assigned the triplets  $\{S_1,7,4\}$ and $\{S_3,3,8\}$.\label{fig:reorgaa}}
		[.4\textwidth]{\includegraphics[width=.4\textwidth]{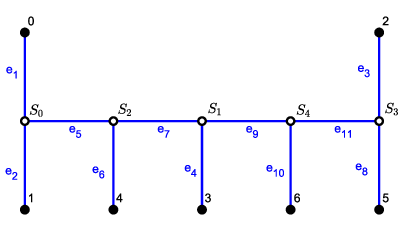}}\\
	\subcaptionbox{The topology $T=(3,5,3,9,11,7,7)$, descendant of $T_p$. If $S_4$ and $S_1$ collide, exchanging $S_5$ for $S_7$ corresponds to the topology $T_a = (3,5,3,7,9,11,15)$. Exchanging it for $3$ leads to $T_b = (3,5,3,4,9,7,7)$.\label{fig:reorgab}}
		[\textwidth]{\includegraphics[width=\textwidth]{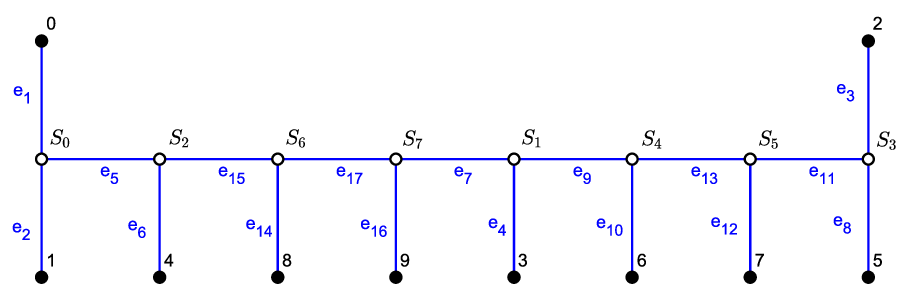}}
	\caption{Schematic representations of Steiner trees and link between adjacent Steiner points.}\label{fig:reorga}
\end{figure}

From now on, we describe how to find the topology corresponding to a reorganized tree. 
In the following, $R_{S_i}$ denotes the regular point introduced with $S_i$.
To understand the whole process, it helps to draw a Steiner tree schematically, with straight lines, by placing $S_i$ on the edge $t_i$, and $R_{S_i}$ on an edge perpendicular to this one, such as in \cref{fig:topo,fig:reorga}. 
Let us define the global index of a Steiner point $S_i$ by $i' = N+i$. 
The global index of a regular point is simply its position in the list of regular points. 
If we describe \cref{fig:reorga3} with this notation, by supposing that $N = 9$, the global index of $S_1$ is $10$, that of $S_3$ is $12$. 
$R_{S_1} = 3$, $R_{S_2} = 4$. 
Actually, we have $R_{S_i} = i + 2$, for all $i = 1,2,...,N-2$ (again, we ignore $S_0$ and the regular points of global indexes $0,1,2$). 
In the following, we note the global index of a point $P$ by $g(P)$.

The first step is to understand the systematic way trees are built. 
Following Smith's procedure on topology $T=(t_1,...,t_k$), we apply the following construction rules. When the edge of index $t_i$ with extremities $P_1$ and $P_2$, such that $g(P_1) < g(P_2)$, is split
\begin{itemize}
	\item  $P_1$ is linked to $S_i$ by an edge of index $t_i$,
	\item $P_2$ is joined to $S_i$ by an edge of index $2*i+3$, and
	\item $R_{S_i}$ is linked to $S_i$ by an edge of index $2*i+2$.
\end{itemize}
\Cref{fig:topo} shows the process of building topology $(3,5)$.

Suppose $S_i$ and $S_h$, with $i>h$, come to the same position. 
To keep the systematic way of enumerating the topologies, it is important that $S_i$ exchange the node that is not $R_{S_i}$. 
Exchanging neighbors actually corresponds to moving $S_i$ and the path linking it to $R_{S_i}$ from its current position to the edge joining $S_h$ and its neighbor.

When we insert a Steiner point $S_i$ in the neighborhood of another Steiner point $S'$, there are obviously three possibilities, one for each edge incident to $S'$. 
Therefore, when $S_i$ collides with $S_h$, the possible new values of $t_i$ in the reorganizations are the indexes of the two other edges incident to $S_h$ \textbf{when $S_i$ was inserted}.  
Algorithmically speaking, we create a table as follows: let $P_1$ and $P_2$ be the extremities of the edge we delete to add the Steiner point $S_i$. 
If $P_1$ is a Steiner point, then we create a triplet $\{P_1,e_1,e_2\}$ for $S_i$, $e_1$ and $e_2$ being the indexes of the two edges adjacent to $P_1$ that are not joining it to $S_i$ (see \cref{fig:reorgaa}). 
We do the same for $P_2$ if it is a Steiner point. 
Each insertion of a Steiner point therefore generates one or two triplets. 
Note that since we build trees by inserting Steiner points one by one, $P_1$ and $P_2$ necessarily have Steiner-index lower than $i$. 
When a reorganization occurs between Steiner points $S_h$ and $S_i$, we examine the point with the bigger index, say $S_i$, and identify the triplet starting with the other one, say $P_1 = S_h$. The new $t_i$ is then $e_1$ or $e_2$, according to which neighbor $S_h$ exchanges.

Changing the position of one Steiner point may have an impact on the Steiner points inserted after it.
We therefore need to ``propagate'' this change. 
\paragraph{On $S_i$'s side (the path $S_i - P_2$), additional changes occur only if $g(P_1) < g(P_2)$.}  
By the rules of construction defined earlier, if $g(P_2) < g(P1)$, when $S_i$ is inserted, $t_i$ is the index of the edge between $S_i$ and $P_2$, let us call it $e$. 
If we insert $S_k$, with $k>i$, between $S_i$ and $P_2$, $t_k = t_i = e$ is the index of the edge between $S_k$ and $P_2$ (because $g(P_2) < g(S_i)$). 
The indexes of the edges between $S_i$ and $P_2$ after the insertion of $S_k$ do not depend on $S_i$ anymore. 
Now, suppose that $S_i$ was not inserted between $P_1$ and $P_2$, $S_k$ is still inserted at $t_k = e$, and Steiner points inserted after that, again, do not depend on $S_i$. 
Therefore, no changes are needed for the Steiner points between $S_i$ and $P_2$.

If $g(P_1) < g(P_2)$, $e$ is now between $S_i$ and $P_1$ so when $S_k$ is inserted between $S_i$ and $P_2$, $t_k \neq e$. If $S_i$ is not inserted however, $t_k = e$ so $S_k$ and every Steiner point between $S_i$ and $P_2$ inserted after it may be impacted by the disappearance of $S_i$.

The Steiner point $S_5$ of \cref{fig:reorga} illustrates that point. 
If $S_4$ is not inserted between $S_1$ and $S_3$ in \cref{fig:reorga3}, to have $S_5$ as its current position in \cref{fig:reorgab} (between $S_1$ and $S_3$), it is clear that $t_5$ must be equal to $9$.

\paragraph{On $S_h$'s side, for similar reasons, changes occur to every Steiner point on the path between $S_h$ and the first node with global index less than that of $S_i$ in the direction of the exchanged neighbor.}
Here, following the direction of the neighbor means following the main path, the one that does not bifurcate to any $R_{S_k}, k=1,...,N-2$. 
Note that on $S_i$'s side, we know that we need to stop when $P_2$ is reached (if it exists) but on $S_h$'s side, we do not know a priori the index of the last point of the path we need to analyze. Let $Q$ be this point.

The method is then the following: 
\begin{itemize}
	\item For each Steiner point $S_j$ on the path $S_i-P_2$:
		\begin{enumerate}
			\item if there is no Steiner point with a smaller index on $S_i$'s side, $t_j =$ index of the edge between $S_i$ and $S_h$,
			\item otherwise, if there is no Steiner point with a smaller index on $P_2$'s side, $t_j = 2 * k + 3$,
		\end{enumerate}
		where $k$ is the index of the smallest Steiner point bigger than $P_2$ between $S_i$ and $S_j$.

	\item For each Steiner point $S_j$ on the path $S_h-Q$:
		\begin{enumerate}
			\item if there is no Steiner point with a smaller index on $Q$'s side, $t_j = 2*i+3$,
			\item otherwise, if there is no Steiner point with a smaller index on $S_h$'s side, $t_j = 2*k+3$,
		\end{enumerate}
		where $k$ is the index of the smallest Steiner point bigger than $S_h$ between $Q$ and $S_j$
\end{itemize}
This process, coupled with the fact the we do not know $Q$ a priori motivates, again, a recursive algorithm.
Following this method, if $S_4$ collides with $S_1$ during the optimization process of the tree described by topology $T=(3,5,3,9,11,7,7)$ (\cref{fig:reorgab}), the reorganizations are $T_a = (3,5,3,7,9,11,15)$ by exchanging $S_7$ and $T_b = (3,5,3,4,9,7,7)$ by exchanging $3$. 
Note that $t_4$ is replaced either by $7$ or $4$, as explained above and seen in \cref{fig:reorgaa}.

Once we know which topologies may result from reorganization, we must determine whether or not to reorganize and, if so, in which order we will explore the topologies. 
Therefore, we define an \emph{interaction criterion}. 
This criterion could be based on a lower bound of the potential trees, their minimum length after complete optimization, the resultant of the surface tension forces exerted on the colliding Steiner point, or anything that one finds suitable. Depending on this criterion, we can then choose which topology should be explored first. 
If the best topology is the current one, there is no need to change the exploration, but if at least one reorganization is better, we can ``jump'' to it in the \mbox{branch-and-bound} tree. 
We can then continue from this topology (eventually examining other reorganizations on the way), and when we have finished exploring all its descendants, we can continue the exploration of the topology we were at first or jump to the second one if it is a better candidate. 

In this way, we obviously explore or prune all nodes of the \mbox{branch-and-bound} tree, and numerical results will show that we do so faster when using topology reorganizations compare to the original algorithm.
\Cref{OES} describes a recursive algorithm for an enumeration scheme equipped with reorganizations. 
This algorithm satisfies three important conditions:
\begin{itemize}
\item It terminates.
\item Every node is explored or cut.
\item No node is explored more than once.
\end{itemize}

\begin{algorithm}
\renewcommand{\algorithmicrequire}{\textbf{Input:}}
\renewcommand{\algorithmicensure}{\textbf{Output:}}

\caption{Enumeration scheme with topology reorganizations}
\label{OES}

\begin{algorithmic}[1]
\Require Set of regular points.
\Statex\hrulefill

\Statex \textbf{Step 0} $l \gets \infty$
\State \Call{explore}{root node, 1, $topvec$}

\Statex 
\Procedure{explore}{$parentNode$, $level$, $topvec$}
	\For {$x:2*level+1\to 1$}
		\State $topvec[level-1] = x$
		\State $currentNode \gets parentNode.child[x]$
		\If {$currentNode.length != 0$}
			\State \textbf{continue for loop}
		\EndIf
		\State \textbf{build} tree yielded by the topology described in $topvec$
		\State \textbf{optimize} tree \label{lin:ch}

		\If {optimization ends normally}
			\If {$currentNode.length > l$}
				\State \textbf{cut} $currentNode$ subtree
			\ElsIf {$currentNode$ is a leaf}
				\State $l\gets currentNode.length$
			\EndIf
		\Else \Comment reorganization occured
			\State $jump \gets currentNode.reorganization$
			\State $topvec \gets$ topology of reorganization
			\State \Call{explore}{jump,level+1,topvec}
			\If {there is a second reorganization}
				\State $jump \gets jump.reorganization$
				\If {$jump.length > l$}
					\State \textbf{cut} $jump$
				\Else
					\State $topvec \gets$ topology of reorganization
					\State \Call{explore}{jump,level+1,topvec}
				\EndIf
			\EndIf
		\EndIf
	\EndFor
	
	\If {there exist children small enough to be explored}
		\State \textbf{order} good children
		\For {$c: 1 \to$ number of good children}
			\State $currentNode \gets parentNode.child[c]$
			\If {$currentNode.length < l$}
				\State $topvec[level-1] = currentNode.id$
				\State \Call{explore}{currentNode,k+1,topvec}		
			\Else
				\State prune $parentNode$
				\State \textbf{break for loop}
			\EndIf
		\EndFor
	\Else
		\State prune $parentNode$
	\EndIf
\EndProcedure

\hrulefill
\Ensure Steiner minimal tree $S$ for the input set.

\end{algorithmic}

\end{algorithm}

We do not operate any reorganizations on the last level, because we wish to reach leaves by going through topologies that tend to give shorter trees. 
Even though, in some cases, this could save a little time, we consider that a reorganization of a leaf topology is not worth the computational cost. 
If the reorganization does indeed give a small enough Steiner tree, we would reach it anyway by continuing the exploration of the \mbox{branch-and-bound} tree. 
However, if the length of the reorganization is too large and needs to be pruned, there is a high probability that its parent or one ancestor is also too large, so we could prune the whole subtree and save the time required to optimize the leaf topology. 
In practice, the second case occurs much more often than the first case. 

In terms of implementation and space, because we use a recursive algorithm, we do not need to store any additional information to be able to ``jump back'' to the topology we came from. 
We can cut branches once their exploration is over so the theoretical maximum number of nodes in the tree at a time is $1 + 3 + 5 + \cdots + (2N-5) + $number of reorganizations to reach the current leaf. 
In practice, reorganizations occur often enough so this number is actually almost never reached.

\FloatBarrier

\section{Pre-optimization pruning method}
\label{sec:pruning}

\cite{Smith} defines the error figure $E$ of a tree as: 
\[E^2 = \sum_{\substack{i \in S\\ ij \in T\\ ik \in T\\ j \neq k}} pos(2(\vec{x}_j - \vec{x}_i) \cdot (\vec{x}_k - \vec{x}_i) + |\vec{x}_j - \vec{x}_i| \cdot |\vec{x}_k - \vec{x}_i|) \]
with $S$ the set of Steiner points, $T$ the set of edges and $pos(x) \equiv max(x,0)$. In a Steiner tree, all angles are at least $120\degree$ so this error figure is simply the ``angle deviation'' from $120\degree$ for the angles smaller than that. $E$ is used as a convergence criterion for the optimization process: the length of a tree is optimum when $E \ll $Length of the tree.

However, in his program, Smith also uses this error figure in the enumeration scheme as a pre-optimization pruning method by testing
\[ L - E < L^*\]
with $L$ the length of the current (not optimized) tree and $L^*$ the length of the current minimal tree. 
\Cref{appendixA} gives an example that shows that doing so can lead to incorrect solutions. 
Nevertheless, pruning the subtree generated by a topology before optimizing it is a good idea, because this will avoid the whole optimization process, a process that can be computationally expensive compared with the other operations of the algorithm. 
One immediate solution is to compute a lower bound for the specified tree. We choose to draw on the concept of toroidal images presented in \cite{ref2}. 
A \emph{cherry} is a pair of regular points incident to the same Steiner point. 
The computation of a lower bound consists of successively replacing a cherry and its Steiner point by a new temporary regular point, called an \mbox{\emph{equilateral point}}, and joining it to the third neighbor of the deleted Steiner point until only two points remain. The distance between these two points gives a lower bound for the length of the specified Steiner tree. 
The equilateral point is a point forming an equilateral triangle with the two vertices of the cherry. 
In the plane, there are two possibilities, but the equilateral point is the one that does not lie on the same side of the cherry as the third neighbor of the Steiner point. 
This point has some interesting properties for a planar Steiner tree but they have no relevance to the current paper. 
Interested readers can easily find a number of papers on the subject, or could refer to \cite{ref4}. 

In higher dimensions, the set of equilateral points becomes more complicated --- it is actually a $N-2$ dimensional torus, hence the name of toroidal image --- but we still choose the planar equilateral point of the cherry for the computation of the lower bound. 
Indeed, a Steiner point lies in the plane formed by its three neighbors and its position is defined by theirs. 
Thus, we can compute a planar equilateral point as follows: 

Let $d \geq 2$ be an integer. 
Let $\textbf{x}_i \in \mathbb{R}^{d}, i = 1, 2, 3$ be three given points such that $\textbf{a} := \textbf{x}_2 - \textbf{x}_1$ and $\textbf{b} := \textbf{x}_3 - \textbf{x}_1$ are linearly independent. 
Hence, the set $X \in \mathbb{R}^{d}$ defined by \[X := \Span \{\textbf{x}_1 + \lambda_1\textbf{a} + \lambda_2\textbf{b} : \lambda_1, \lambda_2 \in \mathbb{R}\}\] is a 2-dimensional affine linear subset in $\mathbb{R}^{d}$.

We wish to find $\textbf{e} \in X$ such that 
\begin{equation}\textbf{the three points $\textbf{x}_1, \textbf{x}_2, \textbf{e}$ form an equilateral triangle.}\end{equation}
The equilateral point can be written as $\textbf{e} = \textbf{x}_1 + r\textbf{a} + t\textbf{b}, \; r,\; t \in \mathbb{R}$. 
Hence, we need to find the coefficients $r$ and $t$. The conditions implied by (1) are
\[|\textbf{x}_2 - \textbf{x}_1| = |\textbf{e} - \textbf{x}_1|, \hspace*{2cm} (\textbf{x}_2 - \textbf{x}_1) \cdot (\textbf{e} - \textbf{x}_1) = |\textbf{x}_2 - \textbf{x}_1|^{2}\cos 60\degree.\]
Rewriting these conditions, we get 
\[|\textbf{a}|^2 = |r\textbf{a} + t\textbf{b}|^2 = r^2|\textbf{a}|^2 + 2rt\textbf{a} \cdot \textbf{b} + t^2|\textbf{b}|^2, \hspace*{2cm} r|\textbf{a}|^2 + t\textbf{a} \cdot\ \textbf{b} = \dfrac{1}{2}|\textbf{a}|^2\]
or
\[t = \dfrac{|\textbf{a}|^2}{\textbf{a} \cdot \textbf{b}}\left(\dfrac{1}{2} - r\right), \hspace*{2cm} |\textbf{a}|^2 = r^2|\textbf{a}|^2 + 2r|\textbf{a}|^2\left(\dfrac{1}{2} - r\right) + \dfrac{|\textbf{a}|^4|\textbf{b}|^2}{(\textbf{a} \cdot \textbf{b})^2}\left(\dfrac{1}{2} - r\right)^2,\]
in which case
\[1 = r - r^2 + \dfrac{|\textbf{a}|^2|\textbf{b}|^2}{(\textbf{a} \cdot \textbf{b})^2}\left(\dfrac{1}{2} - r\right)^2.\]
Finally, we have
\[r = \dfrac{1}{2} \pm \dfrac{\sqrt{3}(\textbf{a} \cdot \textbf{b})}{2D}, \quad t = \mp \dfrac{\sqrt{3}|\textbf{a}|^2}{2D}, \quad \text{with } D := \sqrt{|\textbf{a}|^2|\textbf{b}|^2 - (\textbf{a} \cdot \textbf{b})^2}.\]
There are two solutions for \textbf{e}. 
We are interested in the one that is ``opposite'' $\textbf{x}_3$, that is the  farthest from $\textbf{x}_3$. 

The equilateral point is all we need to compute the lower bound, but note that, 
because a Steiner point is on the segment joining the equilateral point and 
$\textbf{x}_3$,  once $\textbf{e}$ is determined, the Steiner point $\textbf{s}$ can be obtained by
\[\textbf{s} = \textbf{e} + t(\textbf{x}_3 - \textbf{e}).\]
Let $\textbf{c} = \dfrac{1}{3}(\textbf{x}_1 + \textbf{x}_2 + \textbf{e})$, 
because a Steiner point is located on the circumcircle of the triangle defined by a cherry and its equilateral point, we have 
\[|\textbf{s} - \textbf{c}| = |\textbf{e} - \textbf{c}|,\]
and
\[|\textbf{e} - \textbf{c}|^2 = |\textbf{e} - \textbf{c} + t(\textbf{x}_3 - \textbf{e})|^2 = |\textbf{e} - \textbf{c}|^2 + 2t(\textbf{e} - \textbf{c}) \cdot (\textbf{x}_3 - \textbf{e}) + t^2|\textbf{x}_3 - \textbf{e}|^2.\]
Thus, we find the coefficient
\[t = - \dfrac{2(\textbf{e} - \textbf{c}) \cdot (\textbf{x}_3 - \textbf{e})}{|\textbf{x}_3 - \textbf{e}|^2}.\]

\Cref{lin:ch} of \cref{OES} can then be adapted to not only optimize the specified Steiner tree but to first compute a lower bound of its length and running the optimization process only if this lower bound is not longer than the candidate length actually recorded. 
If the lower bound is greater than the length recorded, we can immediately discard this node and its subtree. 
Obviously, this computation gets more time-saving as the number of nodes of a tree increases. 
One could argue that computing a lower bound for trees with few regular points is a waste of computing time.

\FloatBarrier

\section{Numerical results}
\label{sec:results}

In this section, we explore in what measure topology reorganizations can improve the performance of the algorithm proposed by Smith, modified according to \cref{sec:pruning}. 
The goal of this section is not to compete with the stronger algorithms \cite{Fampa}\cite{Fonseca}\cite{Laarhoven} developed the last decade but to assess that topology reorganizations can and should indeed be used as part of the usual enumeration scheme to prune the branch-and-bound tree more.     
First, we present some results for general instances in two, three and four dimensions, although the performance on planar instances is of least importance 
We take a simple interaction criterion: a lower bound of the reorganization is computed and the corresponding tree is then completely optimized if the bound is satisfying. 
The flaw of this strong criterion is the lack of opportunity to save time on the optimization process. 
We focus on three things: the execution time, the number of topologies explored, and the average number of reorganizations that occur for a given number of points in a given dimension. 
As we explore topologies, two cases occur: the trees corresponding to topologies explored ``normally'' are built from scratch while the ones corresponding to reorganizations are already built. 
This second case does not appear in the original algorithm.
Therefore we choose to indicate the number of lower bounds computed to represent the number of topologies explored.
As we do not compute lower bounds before reaching the first leaf, this choice gives a little advantage to the original algorithm but we consider it negligible and report the results without taking it into account.
Implementations of the algorithms were tested using randomly generated instances whose points lay in $\left[-1, 1\right]^d$ for $d=2,3,4$. 
The results are presented as ratios between the version using topology reorganizations and the original algorithm; 
that is, a ratio of 0.8 denotes an improvement of twenty percent.

\begin{table}	
	\caption{Comparison of performances on randomly generated instances.}\label{tab:results}
	\centering
\begin{footnotesize}
	\begin{subtable}[t]{\textwidth}
		\centering
		
		\begin{tabular}{c|c|c|c}
		Nb of points & CPU ratio & Topology ratio & Avg. reorga. \\
		\hline
		8 & 0.7228 & 0.7261 & 223\\

		9 & 0.7760 & 0.7672 & 1220\\

		10 & 0.7853 & 0.7727 & 6576\\

		11 & 0.7967 & 0.7788 & 39137\\

		12 & 0.7760 & 0.7561 & 158651\\
		\end{tabular}
		\caption{In two dimensions.}\label{tab:plane}
	\end{subtable}
	\newline
	
	\begin{subtable}[t]{\textwidth}
		\centering
		\begin{tabular}{c|c|c|c}
		Nb of points & CPU ratio & Topology ratio & Avg. reorga.\\
		\hline

		8 & 0.8682 & 0.8565 & 405\\

		9 & 0.8544 & 0.8527 & 3219\\

		10 & 0.8304 & 0.8409 & 24102\\

		11 & 0.8005 & 0.8168 & 184047\\

		12 & 0.7656 & 0.7877 & 1412586\\
		\end{tabular}
		\caption{In three dimensions.}\label{tab:space}
	\end{subtable}
	\newline
	
	\begin{subtable}[t]{\textwidth}
		\centering
		\begin{tabular}{c|c|c|c}
		{Nb of points} & CPU ratio & Topology ratio & Avg. reorga.\\
		\hline

		8 & 0.8935 & 0.9111 & 528\\

		9 & 0.8589 & 0.8826 & 5078\\

		10 & 0.8115 & 0.8534 & 51611\\

		11 & 0.7712 & 0.8235 & 349401\\

		12 & 0.7526 & 0.8070 & 4074083\\
		\end{tabular}
		\caption{In four dimensions.}\label{tab:4d}
	\end{subtable}
	
\end{footnotesize}
\end{table}

Due to the (possibly very large) difference of time required to solve two different instances with the same number of regular points, we consider benchmarking of random instances relevant if at least several dozens (or hundreds) of instances can be averaged. 
However, the exponential growth of the number of topologies to explored makes the computation of such results for large sets of regular points practically impossible.

The first leaf we reach, that is the first tree including all $N$ regular points, is important in the algorithm, as it becomes the first lower bound of the solution. 
After reaching it, we can start pruning the tree. 
As explained before, Smith's algorithm sorts the child topologies by length when adding to the stack, and the first tree on all regular points produced is the result of a depth first search through the tree. 
This first leaf, or one of the leaves having the same parent, could be of very high quality. 
Therefore, we want to check how the quality of this first bound to the length of the solution is impacted by the use of topology reorganizations. 
More precisely, rather than the very first one, we verify the quality of the bound we obtain after the first exploration of a group of sibling leaves.
On 1000 3-dimensional random instances of 10 regular points, the bound is better in 729 cases using reorganizations. 
However, in the 271 other cases, the difference between the two bounds is on average $3.0000\mathrm{e}{-5}$, whereas the difference when the bound is better is on average $9.529\mathrm{e}{-2}$. 

With the enumeration scheme of the original algorithm, we enumerate the children of a topology, optimize each of them and select the smallest one to go deeper in the \mbox{branch-and-bound} tree. 
With the Steiner tree construction method described in Section 2, we will reach the first minimum in $1+ \sum\limits_{n=1}^{N-4} 2*n + 1 = (N - 3)^2$ steps, where each step denotes moving from one topology to another. 
However, in many cases, two Steiner points will collide during the optimization process. 
Thus, using reorganizations allows to reach (better) leaves faster. 
The minimum number  of steps is $N - 3$, the height of the \mbox{branch-and-bound} tree, but this can only be reached if a jump occurs at every step.

All these results suggest that it is always worth computing the optimization process of the original algorithm with topology reorganizations.

\FloatBarrier

\section{Twin trees}
\label{sec:twintrees}
In this last section, we continue to discuss the relation between topologies.
We unfortunately restrict ourselves to planar problems, but if a generalization to higher dimensions can be found, this may lead to new ways of handling topologies of Steiner trees and of discarding obviously wrong ones.

Let us consider a $2-$dimensional Steiner tree with $S$ and $S'$ two Steiner points of this tree. 
As it was said in \cref{sec:scheme}, by reorganizing their neighborhood, it is possible to get three different topologies $T_1$, $T_2$, and $T_3$. 
Suppose that the neighbors of $S$ are $\{A,B,S'\}$ and the neighbors of $S'$ are $\{C,D,S\}$. 
$A$, $B$, $C$, $D$ may or may not be Steiner points. 
Now, looking at the optimized tree,  if the segments $[AB]$ and $[CD]$ intersect, we call this topology $T_0$ and we show the following:
\begin{theorem}\mbox{ }
\begin{enumerate}
\item{$S$ and $S'$ have the same coordinates,}
\item{the minimal tree representing $T_0$ has length longer or equal to the ones representing the two other possible topologies if we reorganize the neighborhood of $S$ and $S'$,}
\item{the minimal length of the neighborhood of $S$ and $S'$ in $T_0$ is $AB + CD$.}
\end{enumerate}
\end{theorem}
\begin{proof}\mbox{ }
\begin{description}
\item[1, 3]$[AB]$ and $[CD]$ intersect means that $ACBD$ (or $ADBC$) is a convex quadrilateral. 
Suppose that $S'$ is at a different position from $S$. 
Necessarily, one segment incident to $S'$ intersects with one incident to $S$, let say, without loss of generality, that $[AS]$ and $[DS']$ intersect at a point $I$. 
The total contribution of the neighborhood to the length of the tree is $AS + BS + DS' + CS' + SS'$. 
But we have $DI = DS' - S'I$ and $CI < CS' + S'I$ so $DI + CI < DS' + CS'$. 
$I$ lying on $[AS]$, the contribution of $[IS]$ to the length of the tree is $0$, thus we have $DI + CI + IS < DS' + CS' + S'S$, which means that $I$ is a better Steiner point than $S'$, which cannot be since the tree is optimized. 
By the same argument on $S'$, we arrive at the conclusion that $S=S'$ and the contribution of the neighborhood to the length of the optimized tree is equal to $AB + CD$.
\item[2]By \cref{thmSmith}, the optimization process shortens the length of the tree at each iteration. If $S=S'$, looking at the tree, we don't know which topology is represented, because the three of them could be. 
These two facts implies that if a tree is optimized and $S\neq S'$, the topology representing this tree is shorter than one where $S=S'$. 
Otherwise the length $L$ would not be minimum for this tree. 
And if the two Steiner points come to the same position for one of the topology where $[AB]$ and $[CD]$ do not intersect, we obtain the (only one) same tree than the one represented by $T_0$. 
So we have $L_{T_0} \leq L_{T_i}, i=1,2,3.$   
\end{description}
\end{proof}

That being said, we realize that we can actually have a topology where $[AB]$ and $[CD]$ do not intersect reorganizing to a topology where they do and which yields the same tree. 
This motivates us to discuss what we call \emph{twin trees}, that is identical trees generated by different topologies. 
If a topology $T$ yields a tree where two Steiner points collide, if the segments joining their respective neighbors do not intersect, and if it is possible to reorganize it to one where they do, this \emph{other} topology yields the exact same tree. 
Finding this topology follows the process described in \cref{sec:topreorga}. 
We just need to be careful to choose the one where $[AB]$ and $[CD]$ intersect if there is any (there may be none).

This leads to an interesting consideration and a possible new way of pruning the \mbox{branch-and-bound} tree. 
Every time we find a topology $T$ that is not small enough, we cut its descendants. 
But now, we know that if $[AB]$ and $[CD]$ do not intersect for $T$ and it is possible to reorganize $T$ to $T_0$ in such a way that they do, $T_0$ is at least as long as $T$, so we can cut its descendants as well, without any computation of lower bound or optimization. 
We know that if $T$ has colliding Steiner points and is a suitable candidate, so is $T_0$.
And reciprocally, if we explore $T_0$ first and it is suitable, so are its two reorganizations.
This way, we can theoretically avoid optimization of many topologies by simply looking at the neighborhood of adjacent Steiner points.

\FloatBarrier

\section{Conclusion}
\label{sec:conclusion}
 
We have presented a method to reorganize topologies and a systematic way of determining which topology a reorganization corresponds to.
We also showed that when topology reorganizations are used when Steiner points come to the same position, the standard enumeration scheme is improved.
This new scheme allows jumps from a node of the \mbox{branch-and-bound} tree to others of same depth
The algorithm finds better minima faster, meaning that more topologies can be eliminated during the exploration. 
To correct a mistake in the original program, we found a lower bound for a specified tree by computing planar equilateral points, even in dimensions higher than two. 
Although better lower bounds exist, we find the toroidal image concept interesting, and believe that the planar equilateral point might be useful for higher-dimensional problems. 
The different parts of our algorithm are easy to combine and could be merged into more sophisticated procedures.
Finally, we described the notion of twin trees. If this notion could be generalized into higher dimensions, we believe that new, strong, algorithms are at hand.

\appendix 
\section{Erroneous fathoming due to incorrect lower bound}
\label{appendixA}

We present an example to show that the use of the error figure of a tree as a pre-optimization pruning method may not lead to the correct answer. 
This example is made of 10 points in the plane and both results were computed with the same original program proposed by Smith. 
The details are given in \cref{tab:smith}.
We emphasize that if we exchange the positions of the first and last points in the input list, the output also changes.
This should never happen, because the algorithm is independent of the order of the input points. 
In particular, the difference between the two minimal length is not negligible, therefore one proposed solution is obviously wrong.
Running the left example with the pre-optimization pruning turned off, we obtain the solution on the right, confirming that it is indeed the reason of the error.

\begin{table}[H]
\caption{Comparison of two different solutions found for the same input.\label{comparison}}
\centering
\begin{footnotesize}
\begin{tabular}{c|c|c}
\multirow{10}*{\begin{tabular}{@{}c@{}}Input \\ coordinates\end{tabular}} & $(0.61,-0.45)$ & $(0.69,0.41)$\\
& $(-0.83,-0.73)$ & $(-0.83,-0.73)$\\
& $(-0.85,-0.99)$ & $(-0.85,-0.99)$\\
& $(-0.44,0.17)$ & $(-0.44,0.17)$\\
& $(0.18,0.43)$ & $(0.18,0.43)$\\
& $(-0.74,-0.93)$ & $(-0.74,-0.93)$\\
& $(0.09,0.59)$ & $(0.09,0.59)$\\
& $(0.51,-0.87)$ & $(0.51,-0.87)$\\
& $(-0.31,0.70)$ & $(-0.31,0.70)$\\
& $(0.69,0.41)$ & $(0.61,-0.45)$ \\
\hline 
Length & 4.4273061491788859 & 4.178703819479203\\
\hline
Topology& (1,4,3,6,1,11,6) & (1,1,3,7,1,7,12)\\
\hline
Tree & \begin{minipage}{.4\textwidth}
      \includegraphics[width=\linewidth, height=50mm]{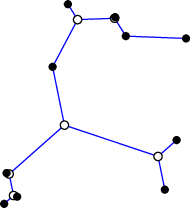}
    \end{minipage}
 & \begin{minipage}{.4\textwidth}
      \includegraphics[width=\linewidth, height=50mm]{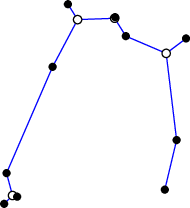}
    \end{minipage}
\label{tab:smith}
\end{tabular}
\end{footnotesize}
\end{table}

\FloatBarrier

\bibliographystyle{plain}
\bibliography{references}

\begin{thebibliography}{10}

\bibitem{ref3}
F.~Chapeau-Blondeau, F.~Janez, and J-L. Ferrier.
\newblock A dynamic adaptive relaxation scheme applied to the {E}uclidean
  {S}teiner minimal tree problem.
\newblock {\em SIAM J. Optimiz.}, 7(4):1037--1053, 1997.

\bibitem{Fampa}
M.~Fampa and K.M. Anstreicher.
\newblock An improved algorithm for computing {S}teiner minimal trees in
  {E}uclidean d-space.
\newblock {\em Discrete Optim.}, 5:530--540, 2008.

\bibitem{review}
M.~Fampa, J.~Lee, and N.~Maculan.
\newblock An overview of exact algorithms for the {E}uclidean {S}teiner tree
  problem in $n$-space.
\newblock {\em Int. T. Oper. Res.}, 23:861--874, 2015.

\bibitem{Fonseca}
R.~Fonseca, M.~Brazil, P.~Winter, and M.~Zachariasen.
\newblock Faster exact algorithm for computing {S}teiner trees in higher
  dimensional {E}uclidean spaces.
\newblock {\em 11th DIMACS Implementation Challenge Workshop}, 2014.

\bibitem{Garey}
M.~R. Garey, R.~L. Graham, and D.~S. Johnson.
\newblock The complexity of computing {S}teiner minimal trees.
\newblock {\em SIAM J. Appl. Math.}, 32:835--859, 1977.

\bibitem{Gilbert}
E.~N. Gilbert and H.~O. Pollak.
\newblock Steiner minimal trees.
\newblock {\em SIAM J. Appl. Math.}, 16(1):1--29, 1968.

\bibitem{ref4}
F.K. Hwang, D.S. Richards, and W~Winter.
\newblock {\em The {S}teiner tree problem}, volume~53 of {\em Ann. Discrete
  Math.}
\newblock Elsevier, Amsterdam, 1992.

\bibitem{ref2}
J.H. Rubinstein, J.~Weng, and N.~Wormald.
\newblock Approximations and lower bounds for the length of minimal {E}uclidean
  {S}teiner trees.
\newblock {\em J. Global Optim.}, 35(4):573--592, 2006.

\bibitem{Smith}
W.~D. Smith.
\newblock How to find {S}teiner minimal trees in {E}uclidean d-space.
\newblock {\em Algorithmica}, 7(1-6):137--177, 1992.

\bibitem{Laarhoven}
J.W. Van~Laarhoven and K.M. Anstreicher.
\newblock Geometric conditions for {E}uclidean {S}teiner trees in
  $\mathbb{R}^d$.
\newblock {\em Comp. Geom.}, 46:520--531, 2013.

\end{thebibliography}

\end{document}